\documentclass[sigconf]{acmart}
\usepackage{booktabs} 

\usepackage[T1]{fontenc}
\usepackage{amsfonts} 
\usepackage{amsmath}
\usepackage{graphicx}
\usepackage{verbatim}
\usepackage{fancyvrb}
\usepackage{url}
\usepackage{xcolor}
\usepackage{soul}

\newcommand{\R}[1][]{\mathbb{R}^{#1}}
\newcommand{\N}[1][]{\mathbb{N}^{#1}}
\newcommand\kdtree[1][\ ]{$k$-d tree{#1}}
\newcommand\ssth[1][\ ]{\textsuperscript{th}{#1}}
\newcommand{\kth}[1][\ ]{$k$\ssth{#1}}
\newcommand{\leqb}{\leq_?}
\newcommand{\ltb}{<_?}

\newcommand\doubleplus{\ensuremath{\mathbin{+\mkern-7mu+}}}
\newcommand\doubleleq{\ensuremath{\mathbin{\leq\mkern-7mu\leq}}}

\RecustomVerbatimEnvironment{Verbatim}{Verbatim}{fontsize=\small}

\AtBeginDocument{%
  }

\setcopyright{acmlicensed}

\acmDOI{xx.xxx/xxx_x}

\acmISBN{xyz}

\acmConference[ ]{ }{ }{ }
\acmYear{2024}
\copyrightyear{2024}

\title{(Nearest) Neighbors You Can Rely On: Formally Verified
    $k$-d Tree Construction and Search in \textsc{Coq}}
\author{Nadeem Abdul Hamid}
\email{nadeem@acm.org}
\orcid{0009-0004-0927-4355}
\affiliation{%
  \institution{Berry College}
  \city{Mount Berry}
  \state{Georgia}
  \country{USA}
  \postcode{30149}
}



\begin{document}

\begin{abstract}
The \kdtree[~]\cite{Bentley75,friedman77} is a classic binary space-partitioning tree used to organize points in $k$-dimensional space. While used in computational geometry and graphics, the data structure  has a long history of application in nearest neighbor search. The objective of the nearest neighbor search problem is to efficiently find the closest point(s) to a given query point, and is the basis, in turn, of common machine learning techniques. We present in this paper a case study in the certified implementation, using the Coq proof assistant, of \kdtree construction from a set of data and the accompanying $K$-nearest neighbors search algorithm. Our experience demonstrates an intuitive method for specifying properties of these algorithms using the notion of list permutations. 
\end{abstract}

\maketitle

\section{Introduction}
\label{sec:intro}

The problem of finding the closest object(s) to a set of data points in 
a metric space has a long history in computing and a wide variety of applications. Formally, the nearest neighbor (NN) search problem  is defined as: for some data set of points in $k$ dimensions, $S \subset \R[k]$, given any  query point $q \in \R[k]$, find a point $p \in S$ that minimizes a distance metric, $\delta_q(p)$. Knuth~\cite{Knuth1973} calls this the \emph{post office problem}, referring to the context of assigning a residence to the nearest post office. The problem is easily generalizable to seeking out the $K$ closest points to $q$ from the data set. Applications of (K)NN search arise in numerous  areas, including machine learning and classification~\cite{Cunningham2021}, recommendation systems~\cite{Kato10}, computer vision~\cite{Turaga10}, and  image retrieval~\cite{Giacinto07}. Wu et al.~\cite{wu2007} include $K$-nearest neighbor classification among the 10 most influential algorithms in the data mining research community.

To improve on the brute-force approach of searching through the entire data set for every query point, effective solutions to the KNN search problem rely on preprocessing the set of points
to enable
more efficient computation of solutions for multiple arbitrary query points. A classic algorithm presented in~\cite{Bentley75,friedman77} introduced the \kdtree[,] a space-partitioning binary tree structure that enables a branch-and-bound search technique, resulting in sub-linear search complexity.

\subsubsection*{Contributions}

In this paper we present a formal verification of \kdtree construction and search algorithms using the Coq proof assistant~\cite{coq2022}. The significant contributions of this work are \emph{formally verified} implementations in Coq of:

\begin{itemize}
    \item A partitioning variant of the classic \emph{quickselect} algorithm, used for finding medians in the data during tree construction;
    \item An algorithm for building a traditional \kdtree structure and searching for the nearest neighbor to a query point; and
    \item A generalized algorithm for searching $K$-nearest neighbors of a query point, using a \kdtree and bounded priority queue.
\end{itemize}

\subsubsection*{Context and Motivations}

As society becomes ever more dependent on applications that rely on core algorithms such as nearest neighbor search, it is worthwhile to move in the direction of applying formal verification methods to such algorithms. Safety-critical applications such as autonomous driving, medical imaging, and cybersecurity systems depend not only on low-level safety properties, but on high-level characteristics of correctness and accuracy. In particular, machine learning-based systems are currently implemented primarily as black boxes with only empirical tests of performance and accuracy.

The results described here are obtained in pursuit of the application of formal methods to specify and verify properties of machine learning systems and algorithms, in particular, classification and regression based on KNN search. This paper focuses on core classical algorithms and their formal verification. Our larger motivation is the development of a top-to-bottom verified implementation of a realistic machine learning system. This necessitates working at different levels of abstraction, reusing and extending existing verified components, and identifying needs and opportunities for new proof automation tools and techniques. Section~\ref{sec:conclusion} situates the result of this paper in the context of future work.

\subsubsection*{Outline} 

In the sections that follow, we provide an overview of \kdtree[s], the nearest neighbor search algorithm, and the Coq proof assistant. Following that we present our implementations in Coq (Section~\ref{sec:impl}) and then explain their formal verification in Section~\ref{sec:formal-verification}. Related work is reviewed in Section~\ref{sec:related} followed by a discussion of future directions and conclusion. An appendix listing Coq statements of lemmas and theorems is also included.

\section{Background}
\label{sec:background}

\subsection{Nearest Neighbor Search Using \kdtree[s]}
\label{sec:background-kdtree}

A \kdtree is a binary tree whose nodes are $k$-dimensional data points. Each level in the tree is associated with one of the $k$ dimensions, usually cycling through them in order, $0 \ldots (k-1)$. Internal (non-leaf) nodes partition the set of nodes in their subtrees based on the dimension axis associated with the level they appear at. In geometric terms, each non-leaf node of a \kdtree splits the $k$-dimensional space along a hyperplane perpendicular to the associated dimension's axis.

For the sake of intuition and ease of explanation, assume our data points are two-dimensional vectors in $\R[2]$. Thus, the root node will partition nodes based on their ``x'' coordinates: all nodes with an x-coordinate less than that of the root will be in the left subtree; the remainder will be in the right. At the next level, the root's immediate children will partition the remaining nodes based on their ``y'' coordinates. Internal nodes at the third level will resume partitioning based on the x coordinate values. 

\begin{figure}[t]
    \centering
    \includegraphics[scale=.6]{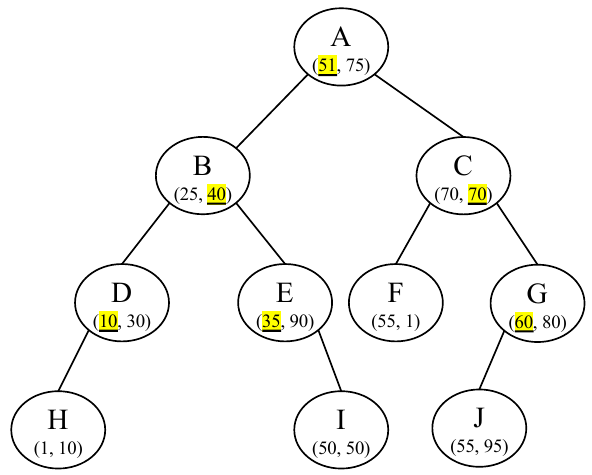}
   \\~\\~\\
    \includegraphics[scale=.6]{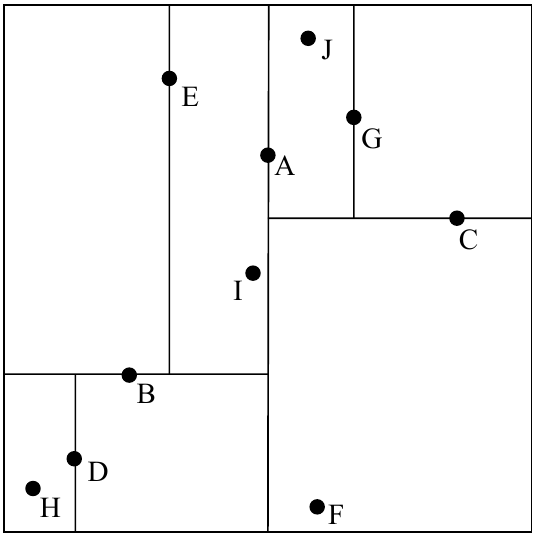}
    \caption{A \kdtree and induced partitions in the $\R[2]$ plane.}
    \label{fig:kdtree-ex}
\end{figure}

A concrete example is provided in Figure~\ref{fig:kdtree-ex}. The bottom portion of the figure illustrates how the structure of the \kdtree on top splits the plane into subplanes, alternating horizontal and vertical divisions. The highlighted coordinate in each node identifies the dimension and value around which the nodes in the subtrees are partitioned. For instance, all the nodes to the right of \textrm{B} have an x-coordinate less than 51 (\textrm{A}'s x-value) and y-coordinate greater than equal to 40 (\textrm{B}'s y-value). 

Constructing a balanced \kdtree is achieved by selecting the data point with the median value for the dimension associated with the current level of the tree. In the example of Figure~\ref{fig:kdtree-ex}, starting with set of points $\{\textrm{A}, \ldots, \textrm{J}\}$, the point A would be selected, for having median x-coordinate value, as the initial root. The rest of the points would be partitioned into two subsets - one with x-coordinates less than 51, and everything else in the other. Each of these two subsets would be recursively processed to build the left and right subtrees of the root. In each case, the point with the median y-coordinate will be selected as the root of the subtree.  Section~\ref{sec:kdtree-construct} describes details of the formalization.

Searching the \kdtree for the nearest neighbor of a query point $q$ proceeds by maintaining the currently known closest point (initially none). The root of the tree is considered and replaces the current closest point if it is closer to $q$. If the root and all points in the (sub)tree are farther away than the closest known point (see discussion on \emph{bounding boxes} in Section~\ref{sec:bbox}), then the entire (sub)tree is discarded (i.e. \emph{pruned}) and the closest known point is maintained as the result. Otherwise, proceed to recursively search the left and right subtrees of the root. In order to prioritize searching the most promising subtree, if the value of the current dimension coordinate of the root is less than the query's, we  search the left subtree first, then the right; otherwise, the right subtree and then the left. 

Generalizing the search to $K$-nearest neighbors\footnote{Throughout the paper, lowercase $k$ denotes the number of dimensions of the data points and uppercase $K$ is the number of nearest neighbors that are sought.} involves maintaining a $K$-bounded \emph{max} priority queue instead of a single closest known point. At each subtree in the search process, if the root is eligible for consideration (i.e. is at least as close to $q$ as the top, or maximum element, of the queue) it is added to the queue. The bounded nature of the queue ensures that if its size exceeds $K$, then the point with \emph{largest} distance from the query will be discarded. While our Coq developments also include verification of the (single) nearest neighbor search algorithm, for the purposes of this paper we focus on the general variant. Section~\ref{sec:knn-search} presents the implementation of $K$-nearest neighbors search in Coq.

\subsection{The Coq Proof Assistant}
\label{sec:coq}

Coq~\cite{coq2022} is an interactive proof assistant based on a higher-order predicate logic extended with inductive data types. By virtue of the Curry-Howard isomorphism relating proofs and programs, the Coq system provides a development environment in which programming and proving are closely intertwined. It enables the user to define data types and functions in a functional programming paradigm while writing rich logical specifications and constructing proofs in the same framework. For brevity, we only highlight here some salient features of Coq.

Inductive definitions allow the introduction of basic data types such as natural numbers\footnote{\url{https://coq.inria.fr/library/Coq.Init.Datatypes.html\#nat}} and polymorphic lists\footnote{\url{https://coq.inria.fr/library/Coq.Init.Datatypes.html\#list}}. Functions that incorporate pattern matching and recursion on inductive objects are defined using \verb+Fixpoint+ and \verb+match+ constructs, as in the following definition of \verb+dropn+:

\begin{Verbatim}
(* produces the list after dropping the first n elements *)
Fixpoint dropn {A:Type} (n:nat) (lst:list A) : list A := 
  match n with 
    | 0 => lst                 (* n = 0 case    *)
    | S n' => match lst with   (* n = 1+n' case *)
                | [] => [] 
                | h :: t => dropn n' t
              end 
  end.
\end{Verbatim}

Recursive definitions must satisfy strict syntactic constraints to ensure termination and totality. Generally speaking, Coq only accepts structurally recursive functions. A common device to define functions with complex recursions is to introduce an extra \verb+nat+ parameter (called the ``fuel'') as the primary recursive argument. A wrapper definition can then be provided for the function that provides enough fuel-- essentially an upper bound on the depth of recursion-- for it to produce a result in every case. Examples of this are in the \verb+quick_select+ and \verb+build_kdtree+ functions presented later.

Besides defining data types, inductive definitions are used in Coq to define logical concepts and propositions. For example, the Coq standard library defines\footnote{\url{https://coq.inria.fr/library/Coq.Sorting.Permutation.html}} \verb|(Permutation lst1 lst2)| as a relation between two lists stating that the elements of one can be reordered to get the other. The specification defines a permutation as being a composition of adjacent transpositions. A \emph{proof} that \verb|lst1| is a reordering of \verb|lst2| involves \emph{constructing} a well-typed object of type  \verb|(Permutation lst1 lst2)|. 

In general, theorems and lemmas in Coq are defined as objects with a type that represents a logical proposition. While proofs may be constructed directly by writing a complete well-formed expression with the correct type, more often they are built interactively and incrementally using a script of proof \emph{tactics}. 
In the remainder of the paper, we write theorem statements in a stylized form and omit proof scripts (which only provide insight anyway when executed interactively). The Appendix provides the precise Coq syntax for the theorem statements presented in the body of this paper. The complete code development, using Coq version 8.16.1, is available at \url{https://github.com/nadeemabdulhamid/knn-search-coq}.

\section{Coq Implementation: Algorithms and Data Structures}
\label{sec:impl}

Efficient search of a \kdtree tree for nearest neighbor queries relies partly on having constructed a well-balanced tree. This in turn, depends on being able to select median values for the root of each subtree. In the following sections, we present our formalized \emph{quickselect} algorithm, data structures for representing bounding boxes, and then the \kdtree construction and search algorithms.

\subsection{Quickselect}
\label{sec:quickselect}

We use a functional variant of Hoare's classic \textsc{Find} algorithm~\cite{hoare-alg65} that selects the $k$th smallest element in a list, in expected linear time. Figure~\ref{fig:quickselect} presents our Coq implementation, which returns not only the $k$th smallest element (where $k=0$ is the smallest of all), but also produces partitions of the remainder of the list. For example,
\verb|(quick_select 3 [15; 90; 32; 7; 86] Nat.leb)| produces 
\verb|  Some ([32; 7; 15], 86, [90])|.

The \verb|steps| parameter of \verb|qsp| is a fuel argument (see Section~\ref{sec:coq}). The \verb|quick_select| function passes the length of  list \verb|l| as the initial fuel value, along with an arbitrary ``less than equal'' comparison predicate \verb|le| for the elements in the list. Lemma~\ref{prop:qsp_works} later establishes that \verb|steps| will never reach 0 in \verb|qsp|. Since Coq functions must be total, we define the workhorse function, \verb|qsp|, to return an \emph{option} type: either \verb|None|, or \verb|Some (a, x, b)| triple containing the $k$th smallest element \verb|x|, along with a list \verb|a| of the $k$ items preceding \verb|x| in \verb|l| were it to be sorted, and the remaining elements \verb|b| after it. Finally, we define a \verb|median_partition| function to partition a given list around its pivot.

For simplicity, our implementation always chooses the head of the list \verb|l| as the \verb|pivot|, then filters the tail, \verb|lst|, into three partitions (line 8) -- elements strictly less than, equal to, and strictly greater than \verb|pivot|. The lengths of these three sublists are examined to determine in which one the \kth smallest element falls. If it is in the equal partition (line 19), then the \verb|pivot| \emph{is} the \kth smallest element (and we divide the partition of equal elements around it to obtain the $k$ elements preceding it when combined with the smaller). Otherwise, the function recursively descends (lines 10, 15) into the appropriate partition, adjusting the value of \verb|k| relative to the partition lengths (line 15).

\begin{figure}[t]
\begin{Verbatim}[numbers=left,fontsize=\scriptsize]
Fixpoint qsp {X:Set} (steps:nat) (k:nat) (l:list X) (le:X -> X -> bool) 
  : option (list X * X * list X) :=
  match steps with 
  | 0 => None (* case should never happen *)
  | S steps' =>
    match l with 
    | nil => None
    | pivot::lst => let '(sm, eq, lg) := partition_sm_eq_lg pivot lst le in
        if k <? length sm then
          match qsp steps' k sm le with
          | None => None
          | Some (a, x, b) => Some (a, x,  b ++ ((pivot::eq) ++ lg))
          end
        else if (length sm + length eq ) <? k then
          match qsp steps' (k - S (length sm + length eq)) lg le with
          | None => None
          | Some (a, x, b) => Some (a ++ (sm ++ (pivot::eq)), x, b)
          end
        else Some (sm ++ (firstn (k - length sm) eq), 
                    pivot, (dropn (k - length sm) eq) ++ lg)
    end end.

Definition quick_select {X} k l le := @qsp X (length l) k l le.

Definition median_partition {X:Set} (l:list X) (le:X -> X -> bool) 
  := quick_select (div2 (length l)) l le.
\end{Verbatim}
    \caption{An expected linear time \kth[]-smallest element selection algorithm.}
    \label{fig:quickselect}
\end{figure}

\subsection{\kdtree Construction}
\label{sec:kdtree-construct}

We define a simple structure representing \kdtree[s], which are either empty or a node with a data point and two subtrees. The natural number in the \verb|node| is the index of the dimension along which data in the subtrees is split by the node. The length of every \verb|datapt| should match the number of dimensions, $k$.  

\begin{Verbatim}
Definition datapt := list nat.

Inductive kdtree :=
| mt_tree : kdtree
| node : nat -> datapt -> kdtree -> kdtree -> kdtree.    
\end{Verbatim}

\begin{figure}[t]
\begin{Verbatim}[numbers=left,fontsize=\scriptsize]
Fixpoint build_kdtree_ (fuel:nat) (k:nat) (data:list datapt) (depth_mod:nat)
  : kdtree :=
  match fuel with
  | 0 => mt_tree (* case should never happen *)
  | S fuel' => 
      match data with
      | [] => mt_tree
      | _ => match median_partition data (ith_leb depth_mod) with
             | None => mt_tree
             | Some (sm, pvt, lg) =>
                 node depth_mod pvt 
                     (build_kdtree_ fuel' k sm (next_depth k depth_mod))
                     (build_kdtree_ fuel' k lg (next_depth k depth_mod))
  end end end.
    
Definition build_kdtree (k:nat) (data:list datapt) : kdtree 
  := build_kdtree_ (length data) k data 0.
\end{Verbatim}
    \caption{Building a \kdtree[.]}
    \label{fig:buildkdtree}
\end{figure}

The \verb|build_kdtree_| function in Figure~\ref{fig:buildkdtree} proceeds by partitioning the \verb|data| list around the median coordinate value of the current depth in the tree. It then builds a node with the pivot point and recursively builds left and right subtrees with the smaller and larger partitions around the pivot. At a level $d$ deep in the tree, \verb|(next_depth k depth_mod)| $ = ((d+1) \bmod k)$; thus, the \verb|depth_mod| argument cycles through the dimensions along which data is split as the recursion descends. \verb|(ith_leb i)| produces a function that compares two data points on their $i^{\rm th}$ dimension. For example, \verb|(build_kdtree 2 [[1;10]; [50;50]; ...; [70;70]])| produces the tree of Figure~\ref{fig:kdtree-ex},
\begin{Verbatim}
    node 0 [51; 75]
      (node 1 [25; 40]
        (node 0 [10; 30] (node 1 [1; 10] ...))
        (node 0 [50; 50] (node 1 [35; 90] ...)))
      (node 1 [60; 80]
        (node 0 [70; 70] (node 1 [55; 1] ...))
        (node 0 [55; 95] ...))    
    \end{Verbatim}

\subsection{Bounding Boxes and Distance Metric}
\label{sec:bbox}

Every node in the \kdtree can be associated with a \emph{bounding box} that describes the minimum and maximum extent of the tree rooted at that node. Bounding boxes can be described as a pair of data points, $[(s_0, \ldots, s_{k-1}), (t_0, \ldots, t_{k-1})]$, where $s_i, t_i \in \N \cup \{ \pm\infty \}$. A point $p = (p_0, \ldots, p_{k-1})$ is contained in such a bounding box if $s_i <= p_i <= t_i$. In the example of Figure~\ref{fig:kdtree-ex}, the bounding box of the root, A, would be $[(-\infty, -\infty), (+\infty, +\infty)]$, while the bounds associated with I would be $[(35, 40), (51, +\infty)]$. In other words, any data point occurring in a subtree rooted at I would have x-coordinate between 35 and 51, and y-coordinate greater than 40 (and unbounded on top). The correspondence to the divided plane in bottom of the figure should hopefully be clear.

In the Coq development, we define a bounding box as a structure containing two lists of \emph{optional} values, where \verb|(Some n)| represents a concrete number $n$, and \verb|None| represents $\pm\infty$.

\begin{Verbatim}[fontsize=\footnotesize]
Inductive bbox : Set 
  := mk_bbox : (list (option nat)) -> (list (option nat)) -> bbox.
\end{Verbatim}

To compute bounding boxes for child nodes, we use a function that splits one bounding box into two, based on a dimension index and a value. 
For example, 
\[\begin{array}{l}
{\tt bb\_split}([(10, \colorbox{yellow}{$-\infty$}, 4), (\infty, \colorbox{yellow}{$30$}, 20)], 1, \colorbox{green}{$11$}) \\
= (~[(10, \colorbox{yellow}{$-\infty$}, 4), (\infty, \colorbox{green}{$11$}, 20)], \\
\qquad[(10, \colorbox{green}{$11$}, 4), (\infty, \colorbox{yellow}{$30$}, 20)]~)
\end{array}\]

Given any arbitrary point $q$ in $k$-dimensional space and a bounding box $B$, an important operation will be to find the point $p \in B$ that minimizes the distance to $q$. (Note, if $q \in B$, then $p = q$.) While not immediately intuitively obvious, the following computes the closest enclosed point:
\[
    {\rm cep}(q, [s, t]) = p \quad {\rm where} \quad p_i = {\rm median}(q_i, s_i, t_i)\footnote{${\it median}(x, y, z)$ is the median of the three values $x$, $y$, and $z$.}
\]

For the purpose of our formalization, we adopt the Manhattan distance metric, $\delta_q(p) = \sum_{i}{|q_i - p_i|}$, throughout our development; though any well-behaved metric would work. Our distance metric $\delta_q(p)$ is thus defined as a Coq function,
\[\verb+sum_dist : datapt -> datapt -> nat+.\]

\subsection{Nearest Neighbor Search}
\label{sec:knn-search}

The generalized KNN search algorithm uses a priority queue to collect candidates for the $K$-nearest neighbors. We adapt the priority queue ADT described in \emph{Verified Functional Algorithms} (VFA)~\cite{appel-vfa} for our purposes. The VFA implementation uses the Coq module system to separate an abstract signature of the queue and its operations (and their properties) from the implementation. Both unsorted list and traditional binary heap implementations are developed in VFA. We generalize the priority queue to contain any type of data elements and equip it with a \verb|key| function mapping data elements to a comparable type (like \verb|nat|). Additionally, we added \verb|peek| and \verb|size| operations to the ADT. Like \verb|delete_max|, \verb|peek| produces \verb|None| if applied to an empty queue. Since this is a functional implementation, \verb|delete_max| produces an updated queue along with the removed element. Figure~\ref{fig:priqueue} also presents a bounded insert operation that is used to maintain a queue with maximum size $K$.

\begin{figure}[t]
\begin{Verbatim}[numbers=left,fontsize=\scriptsize]
Parameter priqueue: forall A:Type, (A -> t) -> Type.

Parameter empty: forall A key, priqueue A key.
Parameter insert: forall A key, A -> priqueue A key -> priqueue A key.
Parameter delete_max: forall A key, priqueue A key -> option (A * priqueue A key).
Parameter peek_max: forall A key, priqueue A key -> option A.
Parameter size: forall A key, priqueue A key -> nat.

Definition insert_bounded (K:nat) A key (e:A) (pq:priqueue A key) 
  : priqueue A key :=
  let updpq := (insert A key e pq)
  in if K <? (size A key updpq) 
     then match delete_max _ key updpq with 
          | None => updpq               (* should never happen *)
          | Some (_ , updpq') => updpq' (* the removed element is discarded *)
          end
     else updpq.
\end{Verbatim}

    \caption{The priority queue ADT and bounded insert operation.}
    \label{fig:priqueue}
\end{figure}

\begin{figure}[t]
\begin{Verbatim}[numbers=left,fontsize=\scriptsize]
Fixpoint knn (K:nat) (k:nat) (tree:kdtree) (bb:bbox) (query:datapt) 
             (pq:priqueue datapt (sum_dist query)) 
  : priqueue datapt (sum_dist query) :=
  match tree with 
    | mt_tree => pq
    | node ax pt lft rgt =>
        let body (pq':priqueue datapt (sum_dist query)) := 
            let dx := nth ax pt 0 in
            let bbs := bb_split bb ax dx in
            if (ith_leb ax pt query) 
            then (knn K k rgt (snd bbs) query
                    (knn K k lft (fst bbs) query pq'))
            else (knn K k lft (fst bbs) query
                    (knn K k rgt (snd bbs) query pq')) 
        in
        match (peek_max _ _ pq) with
        | None => body (insert_bounded K _ _ pt pq)
        | Some top => if (K <=? (size _ _ pq))
                && ((sum_dist query top) 
                        <? (sum_dist query (closest_edge_point query bb)))
                    then pq
                    else body (insert_bounded K _ _ pt pq)
        end
  end.

Definition knn_search (K:nat) (k:nat) (tree:kdtree) (query:datapt) 
  : list datapt :=
  pq_to_list 
    (knn K k tree 
         (mk_bbox (repeat None k) (repeat None k)) query 
         (empty datapt (sum_dist query))).
\end{Verbatim}

    \caption{The $K$-nearest neighbors search algorithm.}
    \label{fig:knn-search}
\end{figure}

Finally, Figure~\ref{fig:knn-search} presents the complete KNN search algorithm. Recall, $K$ is the number of nearest neighbors desired, while $k$ is the number of dimensions. The main function, \verb|knn_search| (line 26), initializes an empty priority queue (using the distance from the query to compute key values) and an unconstrained bounding box, and handles converting the final returned priority queue into a list. In the \verb|knn| function, as long as the \verb|tree| is not empty, we first attempt to peek at the top of the queue (line 16). If the priority queue is full (line 18) and the closest point to the query that lies in the current bounding box is farther away than the element at the top of the queue, then the entire tree is ignored and the queue produced unchanged (line 21). Otherwise, the root node is inserted into the queue (lines 17, 22) and the \verb|body| is executed. In the \verb|body| expression, \verb|dx| is the coordinate value of the tree root for the axis (dimension index) around which the tree is partitioned at this level. The current bounding box is split into its two components, \verb|bbs|, and then two recursive calls are made according to the most promising order, as described in Section~\ref{sec:background-kdtree}.

\section{Formal Verification}
\label{sec:formal-verification}

We now highlight aspects of our formal specification and verification of the algorithms and data structures defined in the preceding section. While presented here in top-down order, the appendix contains Coq statements of supporting lemmas and intermediate definitions in the order they appear in the actual proof scripts. Notations used here include $|l|$ for length or size, $(l1 \doubleplus l2)$ for list concatenation, and $(l1 \bowtie l2)$ for \verb+(Permutation l1 l2)+. The binary relation on lists, \texttt{all\_in\_leb}, denoted by $l_1 \doubleleq l_2$, holds if $\forall e_1 \in l_1, e_2 \in l_2, e_1 \leqb e_2$. The  $\leqb$ relation is a total order (see Theorem~\ref{prop:quick_select_exists_correct}).


\subsection{KNN Search}

To specify correctness of the overall KNN search algorithm, suppose we have a list of \verb+data+ points from which we build a $k$-d \verb+tree+. Then, given a \verb+query+ point, we expect the \verb+result+ of \verb+knn_search+ from Figure~\ref{fig:knn-search} to satisfy the following:
\begin{enumerate}
    \item The \verb+result+ should be a list of $K$ elements, but no more than $|$\verb+data+$|$.
    \item The \verb+result+ should be a subset of the original data. We express this by positing that there exists a \verb+leftover+ list which, together with \verb+result+, is some permutation of \verb+data+ (effectively describing a reordered partitioning).
    \item Every element in the \verb+result+ should be, according to the metric of \emph{distance from the query point}, less than or equal to  every other element from the \verb+data+ that is not in the \verb+result+ (i.e. that is in \verb+leftover+). In other words, they are the elements with the smallest distances from the \verb+query+.
\end{enumerate}

\noindent Establishing these properties depends on preconditions captured in the full Coq statement of what is ultimately the final theorem of our development:

\begin{theorem}[knn\_search\_build\_kdtree\_correct]\label{prop:knn_search_build_kdtree_correct}

\begin{Verbatim}

forall (K:nat) (k : nat) (data : list datapt),
  0 < K ->      (* at least one nearest neighbor sought *)
  0 < length data ->                 (* data is non-empty *)
  0 < k ->                (* dimension space is non-empty *)
  (forall v' : datapt,    (* all data points well-formed, *)
        In v' data -> length v' = k) ->   (* k dimensions *)
  forall tree query result,
      tree = (build_kdtree k data) ->
      result = knn_search K k tree query ->
      exists leftover, 
        length result = min K (length data)            (*1*)
        /\ Permutation data (result ++ leftover)       (*2*)
        /\ all_in_leb (sum_dist query) result leftover.(*3*)
    \end{Verbatim}   
\end{theorem}

In our experience, it is easier to separate out and prove the first property independently of the other two. The following generalized proposition is established by induction on $t$, proceeding by case analysis on the possible execution paths of \verb+knn+ (Figure~\ref{fig:knn-search}). Its proof relies on correctness of the bounded insertion operation (\verb+insert_bounded+) with respect to maintaining a maximum queue size of $K$. \verb|Contents_kdtree| is an inductively-defined relation that associates a \kdtree with a list of all the data points it contains in its nodes.\footnote{
As noted in Section~\ref{sec:knn-search}, we adapt the VFA priority queue module for our use. That library frames correctness of the ADT using a model-based specification: the abstract \texttt{priqueue} type is associated with a \emph{representation invariant} (\texttt{priq}) as well as an \emph{abstraction relation} (\texttt{Abs}) that establishes a correspondence between the \texttt{priqueue} type and the concrete \texttt{list} type. This enables reasoning about programs that use a \texttt{priqueue} through simpler reasoning about behaviors of lists, regardless of how the \texttt{priqueue} is implemented. The proposition statements in this section elide hypotheses involving these invariants, but they appear in the full Coq statements in the appendix.}

\begin{lemma}[knn\_search\_build\_size\_gen]\label{prop:knn_search_build_size_gen}
    For any data point list $D$ and tree $t$, where ${\rm Contents\_kdtree}(t) = D$; and priority queue $pq$, where $|pq| \leq K$,
\[|{\rm knn}(K, k, t, B, q, pq)| = \min(K, |D| + |pq|).\]
\end{lemma}

In many cases, propositions like Lemma~\ref{prop:knn_search_build_size_gen} come in pairs, with another top-level lemma (named \textsc{knn\_search\_build\_size} in this case) that is specialized to initial conditions; for example, that $t = {\rm build\_kdtree}(k, D)$ and $B = [(-\infty, -\infty), (+\infty, +\infty)]$. Applying this to the proof of Theorem~\ref{prop:knn_search_build_kdtree_correct} depends on the relationship between \verb|Contents_kdtree| and \verb|build_kdtree| (see Lemma~\ref{prop:build_kdtree_contents_perm_gen}).

For the latter two properties of Theorem~\ref{prop:knn_search_build_kdtree_correct}, we develop another pair of lemmas expressing that \verb|knn_search| effectively partitions the input data into two lists, $R$ (result) and $L$ (leftover), where $R \doubleleq L$. The top-level one is stated:

\begin{lemma}[knn\_full\_relate]\label{prop:knn_full_relate}
    Given a list of data points $D$, query point $q$, and tree $t = {\rm build\_kdtree}(k, D)$,
\[\exists R, L,~~ {\rm knn\_search}(K, k, t, q) = R ~~\wedge~~ D \bowtie (R \doubleplus L) ~~\wedge~~ R \doubleleq L .\]
\end{lemma}

The associated generalized lemma unravels the details of the data flow through the algorithm, and is proved by induction on the structure of the tree:

\begin{lemma}[knn\_full\_relate\_gen]\label{prop:knn_full_relate_gen}
    For any data point list $D$, tree $t$ bounded by $B$ (where ${\rm Contents\_kdtree}(t) = D$), priority queue $pq$, and result $R = {\rm knn}(K, k, t, B, q, pq)$,
    
\[\begin{array}{rl}
    \exists pq_{sm}, pq_{lg}, D_{sm}, D_{lg}, L, &  (pq_{sm} \doubleplus pq_{lg}) \bowtie pq \quad \wedge \\
                                                 &  (D_{sm} \doubleplus D_{lg}) \bowtie D \quad \wedge \\
                                                 &  (pq_{sm} \doubleplus D_{sm}) \bowtie R \quad \wedge \\
                                                 &  (pq_{lg} \doubleplus D_{lg}) \bowtie L \quad \wedge \\
                                                 &  R \doubleleq L \quad \wedge \\
                                                 &  ((|R| < K) \Rightarrow L = []).
\end{array}\]
\end{lemma}

\begin{figure}[t]
    \centering
    \includegraphics[scale=.45]{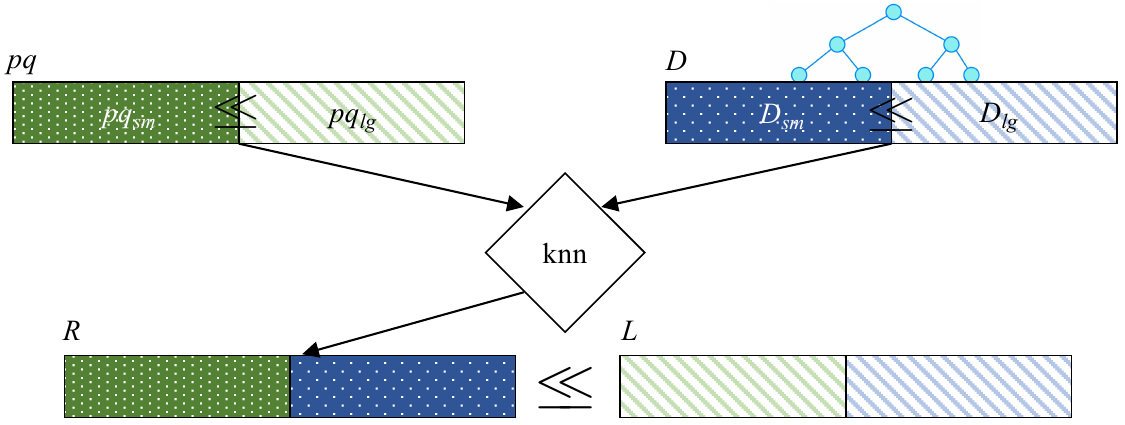}
    \caption{Partitions induced by the \texttt{knn} function.}
    \label{fig:knn-partitions}
\end{figure}

\noindent
A graphical intuition for Lemma~\ref{prop:knn_full_relate_gen} is given in Figure~\ref{fig:knn-partitions}. The effect induced by \verb|knn| on its input priority queue (viewed abstractly as a list) and list of data points (contained in the \kdtree[]) is to partition each of those lists into two chunks, where all elements of the first chunk ($pq_{sm}$ and $D_{sm}$) are less than or equal to those in the second ($pq_{lg}$ and $D_{lg}$, respectively). Then, \verb|knn| selects out the two chunks of lesser elements to form the result it produces, leaving behind a (discarded) result that contains the two larger chunks of each list. While it appears intuitively correct, the proof of this lemma was extremely long and tedious to develop. One reason was the lack of automation for reasoning about \verb|Permutation|s (discussed later). The other was that large portions of the proof tactics were duplicated to cover the different execution paths in the function (the conditional expressions leading to the evaluation of the \verb|body| in Figure~\ref{fig:knn-search}). When reasoning through the nested recursive calls, each intermediate result gets partitioned into additional pieces as it feeds through the next application of \verb|knn|, resulting in invocations of the inductive hypotheses with slightly different contexts.

\subsection{Bounding Boxes and \kdtree[s]}

Lemmas~\ref{prop:knn_full_relate} and~\ref{prop:knn_full_relate_gen} also rely on properties about bounding boxes and their relationship to \kdtree[s]. While not explicitly represented in the \verb|kdtree| data type, bounding boxes are crucial for reasoning about the properties of the search algorithm. We define a computational function, \verb|kdtree_bounded|, to check if a tree is properly constrained by a given bounding box. This does not simply check that every point in the tree occurs in the root's bounding box, but rather the more nuanced property that subtrees are constrained by the boxes appropriately split (Section~\ref{sec:bbox}) according to the dimension and value associated with the root.

We can then establish that the \verb|build_kdtree| algorithm (Figure~\ref{fig:buildkdtree}) constructs a tree that is properly constrained by an initially unbounded box. The proof of this lemma relies on properties of \verb|quick_select| (Theorem~\ref{prop:quick_select_exists_correct}):

\begin{lemma}[build\_kdtree\_bounded]\label{prop:build_kdtree_bounded}
    Given a list of data points, $D$, each of dimension $k$,
    \[{\rm kdtree\_bounded}({\rm build\_kdtree}(k, D), [(-\infty, -\infty), (+\infty, +\infty)]).\]
\end{lemma}

One other crucial proposition related to bounding boxes is that the \verb|cep| function (Section~\ref{sec:bbox}) does in fact compute the closest enclosed point. This is necessary for the proof of Lemma~\ref{prop:knn_full_relate_gen}:

\begin{lemma}[cep\_min\_dist]\label{prop:cep_min_dist}
    Given a point $q$ and bounding box $B$,
$ \forall p \in B, \delta_q({\rm cep}(q, B)) \leq \delta_q(p).$
\end{lemma}

Besides conforming to bounding box constraints, the \verb|build_kdtree| algorithm should result in a tree whose contents are exactly some permutation of the items in the original list. Establishing this also requires validating the viability of the fuel argument for \verb|build_kdtree_| in Figure~\ref{fig:buildkdtree} (because if the fuel ran out early, the function would not process all the items in the list):

\begin{lemma}[build\_kdtree\_contents\_perm\_gen]\label{prop:build_kdtree_contents_perm_gen}
    For a list of data points, $D$, if $k > 0$, $dmod < k$, and $f \geq |D|$, then $\exists l,$ ${\rm Contents\_kdtree}({\rm build\_kdtree\_}(f, k, D, dmod)) = l ~~ \wedge ~~ l \bowtie D$.
\end{lemma}
\noindent Here $dmod$ represents the value $(d \bmod k)$ where $d$ is the current depth level of the tree being constructed through the recursive calls of \verb|build_kdtree_| (see \verb|depth_mod| in Section~\ref{sec:kdtree-construct}). In addition to having a sufficient bound on the fuel $f$, correct behavior depends on the number of dimensions $k$ being non-zero and the value of $dmod$ being a valid dimension index. The proof is fairly straightforward through unfolding the different execution paths in \verb|build_kdtree_|. The {\rm\bf\small (quick\_select\_permutation)} lemma of Theorem~\ref{prop:quick_select_exists_correct} in the next section is necessary for establishing the permutation property.


\subsection{Quickselect}

To account for the fuel argument used by \verb|qsp| (Figure~\ref{fig:quickselect}), we prove that as long as \verb|steps| is at least the length of the input list, and \verb|k| is less than its length, \verb|quick_select| will always produce \emph{some} result (as opposed to \verb|None|). 

\begin{lemma}[qsp\_works]\label{prop:qsp_works}
If $k < |l|$ and $|l| \leq {\rm steps}$, then 
\[\exists p, {\rm qsp}({\rm steps}, k, l, \leqb) = {\rm Some~}p.\]
\end{lemma}

With this established, we can specify and prove properties of \verb|quick_select| as stated by the following theorem. The labels in parentheses correspond to names of lemmas in the Coq proof script, each property stated and proved independently.

\begin{theorem}[quick\_select\_exists\_correct]\label{prop:quick_select_exists_correct} 
    For any well-behaved ``less than or equal'' operation, $\leqb$, if $k < |l|$, then 

\begin{enumerate}
    \item $\exists l_1, v, l_2, \\{\rm quick\_select}(k, l, \leqb) = {\rm Some\ } (l_1, v, l_2) $ {\rm\bf\scriptsize (quick\_select\_always\_some)}
    \item $l \bowtie (l_1 \doubleplus [v] \doubleplus l_2) \quad$ {\rm\bf\scriptsize (quick\_select\_permutation)}
    \item $|l_1| = k \quad$ {\rm\bf\scriptsize (qsp\_position)}
    \item $\forall x \in l_1, x \leqb v  \quad$ {\rm\bf\scriptsize (qsp\_all\_smaller)}
    \item $\forall x \in l_2, v \leqb x  \quad$ {\rm\bf\scriptsize (qsp\_all\_larger)}
\end{enumerate}
\end{theorem}

For \verb|quick_select|, (1) we always get a partition of the original list into three pieces: two lists and an element. The combination of these pieces should be a reordering of the input list (2). If $v$ is meant to be the \kth smallest element (where $k=0$ for the minimum of entire list), then $l_1$ should contain exactly $k$ elements (3). We also expect that everything in $l_1$ is less than or equal to $v$ (4); and symmetrically, that everything in $l_2$ is greater than or equal to $v$ (5). The properties of $\leqb$ needed to establish (4) and (5) above are that it be transitive and total:

\begin{Verbatim}
Definition le_props {X:Set} (le:X -> X -> bool) : Prop :=
    (forall a b c, le a b = true -> le b c = true 
                                 -> le a c = true) /\
    (forall a b, le a b = true \/ le b a = true).
\end{Verbatim}

Theorem~\ref{prop:quick_select_exists_correct} above is one possible formulation of correctness for \verb|quick_select|. Alternatively, an early iteration used the following (in which the function only produced the median value):

\begin{quotation}
\noindent If ${\rm quick\_select}(k, l, \leqb) = {\rm Some\ } v$, then
    \begin{enumerate}
        \item $|({\rm filter} ~ (\_ \ltb v) ~ l)| \leq k$,\footnote{This says, the number of elements in $l$ strictly less than $v$ is less than or equal to $k$.} 
        \item $|({\rm filter} ~ (\_ =_{?} v) ~ l)| \leq |l|$, ~ and
        \item $|({\rm filter} ~ (v \ltb \_) ~ l)| < |l| - k$.
    \end{enumerate}
\end{quotation}

In our assessment, while logically equivalent and more concise, this formulation is much less intuitive. It is also not convenient at all to work with when reasoning about a larger algorithm that uses \verb|quick_select|. Appealing to the concept of \verb|Permutation| enables concise and intuitive specification of the algorithm's behavior.



\section{Related Work}
\label{sec:related}

Verification of classic, or ``textbook,'' algorithms and data structures has a long history in mechanical verification and theorem proving. The subtle and nuanced invariants involved mean that the process often needs to be guided interactively (as opposed to using fully automated methods). Nipkow et al.\cite{nipkow20} reviews the state of the art at the beginning of the current decade. As surveyed in that work, there are a variety of systems and frameworks used for algorithm verification. In our use of Coq, the present work resembles algorithms and data structures verified by Appel~\cite{appel-vfa}, although \kdtree[s] are slightly more complex than the material covered there. In terms of the complexity of the data structure and algorithm, our work is similar to the verification of the Gale-Shapley stable matching algorithm in~\cite{hamid10}.

To our knowledge, there is no prior work in Coq formally verifying \kdtree[s] or any type of nearest neighbor search algorithm. After completing our verification project, we did became aware of an earlier, independent formalization and verification based on the Isabelle/HOL proof assistant, not described or published other than the script in the Archive of Formal Proofs~\cite{afp-kdtree-rau19,afp-mom-eberl17}. The latter verifies a more sophisticated pivot selection algorithm, median-of-medians, whereas our formalization simply picks an arbitrary pivot, since our primary focus was on the KNN search algorithm.  In contrast, our implementation of quickselect produces a balanced partition of the input list in addition to its median value. In Rau's development~\cite{afp-kdtree-rau19}, a different form of \kdtree[s] was implemented, where all data points occur at the leaves of the tree and internal nodes only store the dimension (axis) being split on and a median value for comparison. Thus, their search implementation proceeds through the tree and only inserts into the queue when it reaches the leaves. Our data structure, storing data points at all nodes in the tree, appears closer to traditional implementations. The formulation of the KNN search algorithm verified in this paper also appears more true to standard formulations, such as in the specification of bounding boxes (``geometric boundaries'' of~\cite{friedman77}) and use of a priority queue ADT. 

In terms of specification of the overall algorithms, while appearing logically equivalent to ours (a good thing!), \cite{afp-kdtree-rau19,afp-mom-eberl17} appeal to notions of sorting and sortedness, whereas we avoid any dependence on notions of sorting and formulate our propositions using more elementary properties of \verb|Permutation| and \verb|all_in_leb|. We feel the use of permutations for the specification is simpler and more understandable, and in some cases also provides better clarity and understanding about the properties and data flow of the algorithm (\`{a} la Figure~\ref{fig:knn-partitions}). Finally, there are differences in the underlying logic of the systems themselves, discussion of which is beyond the scope of this paper. For example, function definitions such as \verb|knn_search| in Coq, being actually computational, must satisfy syntactic termination conditions. Thus, a well-typed function is guaranteed to terminate by virtue of its definition being type checked. In Isabelle/HOL, termination is established after the fact (though automatically in many cases) through additional external proof obligations generated by the framework~\cite{krauss11}.

\section{Discussion and Conclusion}
\label{sec:conclusion}

As commonly the case with formalization exercises, our experience with the KNN search algorithm forced a complete understanding of all its operational nuances. In particular, since portions of the development were attended by a team of undergraduate students, correctness of the pruning behavior and understanding of its rigorous justification, instead of simply an intuitive sense, became very necessary. Low-level details that are often omitted, assumed, or ignored in standard presentations had to be explicitly worked out and stated in order for the proof development to go through. In fact, presentations of the \kdtree[-]based KNN search algorithm generally omit rigorous explanation of correctness, assuming that it is self-evident. Thus, in developing the specifications and proofs described above, we followed our own initiative, rather than following a prior published ``paper'' proof.

While achievable with reasonable effort, the formal developments were still non-trivial. In our project, we used only the standard Coq library and little proof automation, leading to some long proof scripts, particularly for Lemma~\ref{prop:knn_full_relate_gen}. Patterns  emerged in the proofs, where many progressed by induction on the \kdtree structure and then case analysis of the three conditional expressions of the algorithm (lines 9, 17, 18 in Figure~\ref{fig:knn-search}). Since there were symmetric, but different orders, of the recursive calls in the \verb|body|, this led to duplicate runs of tactics in the script with slightly different initial contexts. This leaves an opportunity to explore better ways to structure or refactor the proofs to abstract the commonalities based on the execution paths of the algorithm.

An alternate avenue for future exploration would be better proof automation~\cite{cpdt} and the use of more sophisticated techniques or extensions~\cite{ssreflect}. In particular, while we believe the concept of list permutations is a good way to express specifications, the most verbose and tedious portions of the proof involved deriving facts about permutation relationships. Reasoning about permutations between lists seems like a good candidate for proof automation. Many exercises in~\cite{appel-vfa} also lead to tedious proofs involving permutations, but to our knowledge there is no existing tactic library to assist with such.\footnote{The closest we found is~\cite{braibant11}, a general-purpose Coq plugin for rewriting modulo associativity and commutativity with some limited application to lists and permutations.} Developing a decision procedure for this purpose would contribute a useful tool to the Coq ecosystem and we hope to pursue this in the short term.

Coq provides a \verb|Program|~\cite{sozeau07} tactic that would allow defining functions like \verb|qsp| and \verb|build_kdtree_| without a fuel argument. It does, however, introduce its own additional layers of complexity in the verification process. In future work, the trade-offs of using that alternate definition could be explored.

There are several different directions we envision future work could pursue in the long term. One would be to develop a verification of an imperative implementation of the algorithms and data structures described in this paper. While that involves additional levels of complexity (and would necessitate reasoning about a separate language, e.g. C or Java), in developing our work for this paper we attempted to stay as close to traditional presentations of the algorithms as possible. We believe this will make it easier to port the specifications and proofs to an imperative context. There are also a number of alternate, modern space-partitioning tree variants used for both exact and, particularly, \emph{approximate} KNN search. Applying our methods to any of these is a way to make the results more potentially applicable for practical, industry use. 

Finally, we hope to extend our current results to explore specification and verification of machine learning algorithms built upon nearest neighbor search. As alluded to in the introduction, there are a wide variety of areas where such techniques are used that are becoming increasingly safety-critical. We envision formal methods and theorem proving techniques as a useful tool for developing and verifying high-level operational properties of such systems, built upon core data structures and algorithms such as those verified in this paper.

\begin{acks}
The author would like to thank his students, Matthew Bowker, Jessica Herring, and Bernny Velasquez for participating in discussions on this project and preliminary efforts on verification of the single nearest neighbor algorithm. Another team of students, Riley Croker, Britt Smith, and Jeffrey Stephens, worked much earlier on an initial formalization of quick\_select in Coq, which served as a base for the version presented in this work.

The author would also like to thank the anonymous reviewers for their valuable comments and constructive suggestions on earlier versions of the manuscript.
\end{acks}

\bibliographystyle{ACM-Reference-Format}
\bibliography{mybib}

\clearpage
\appendix
\section{Coq Statements of Lemmas and Theorems}

\begin{Verbatim}
(** Quickselect **)

Theorem qsp_works :
  forall (X:Set) steps (k:nat) (l:list X) (le:X -> X -> bool),
  k < length l ->
  length l <= steps ->
  exists p, qsp steps k l le = Some p.

Theorem quick_select_always_some :
  forall (X:Set) (k:nat) (l:list X) (le:X -> X -> bool),
  k < length l ->
  exists l1 v l2, quick_select k l le = Some (l1,v,l2).

Theorem quick_select_permutation :
  forall (X:Set) (k:nat) (l:list X) (le:X -> X -> bool) l1 l2 v,
  quick_select k l le = Some (l1, v, l2) -> 
  Permutation l (l1 ++ v :: l2).

Theorem qsp_position : forall (X:Set) steps (k:nat) (l:list X) 
    (le:X -> X -> bool) l1 l2 v,
  qsp steps k l le = Some (l1, v, l2) -> 
  length l1 = k.

Theorem qsp_all_smaller : forall (X:Set) steps (k:nat) (l:list X) 
    (le:X -> X -> bool) l1 l2 v (Hle:le_props le),
  qsp steps k l le = Some (l1, v, l2) -> 
  forall x, In x l1 -> le x v = true.

Theorem qsp_all_larger : forall (X:Set) steps (k:nat) (l:list X)
    (le:X -> X -> bool) l1 l2 v (Hle:le_props le),
  qsp steps k l le = Some (l1, v, l2) -> 
  forall x, In x l2 -> le v x = true.

Theorem quick_select_exists_correct :
  forall (X:Set) (k:nat) (l:list X) (le:X -> X -> bool),
  le_props le ->
  k < length l ->
  exists l1 v l2, 
    quick_select k l le = Some (l1,v,l2) /\
    Permutation l (l1 ++ v :: l2) /\
    length l1 = k /\
    (forall x, In x l1 -> le x v = true) /\
    (forall x, In x l2 -> le v x = true).

(** Bounding Boxes and k-d Trees **)

Definition bb_split (bb:bbox) (d:nat) (v:nat) : bbox * bbox :=
  let `(mk_bbox bb_min bb_max) := bb in
  ( mk_bbox bb_min (list_set bb_max d (Some v)),
    mk_bbox (list_set bb_min d (Some v)) bb_max ).
    (* list_set updates the d'th coordinate of a data point *)

Fixpoint kdtree_bounded (tree:kdtree) (bb:bbox) : bool :=
    match tree with
    | mt_tree => true
    | node ax pt lft rgt => 
       bb_contains bb pt
       && kdtree_bounded lft (fst (bb_split bb ax (nth ax pt 0)))
       && kdtree_bounded rgt (snd (bb_split bb ax (nth ax pt 0)))
    end.    
Lemma In_kdtree_bounded_bb_contains :
    forall tree bb v, kdtree_bounded tree bb = true -> 
    In_kdtree v tree -> bb_contains bb v = true.

Lemma build_kdtree_bounded : forall (k:nat) (data:list datapt),
    0 < k ->
    (forall v', In v' data -> length v' = k) ->
    kdtree_bounded (build_kdtree k data)
               (mk_bbox (repeat None k) (repeat None k)) = true.

Lemma cep_min_dist :
    forall vs mins maxs ws, 
    cep vs mins maxs = ws -> 
    forall pt, bb_contains (mk_bbox mins maxs) pt = true ->
    sum_dist vs ws <= sum_dist vs pt.

Lemma build_kdtree_contents_perm_gen
    : forall fuel k data depth_mod,
      0 < k ->   (* need for next_depth_lt *)
      depth_mod < k ->
      length data <= fuel) ->
      exists lst,
        Contents_kdtree 
            (build_kdtree_ fuel k data depth_mod) lst /\ 
        Permutation data lst.

Lemma build_kdtree_contents_perm
    : forall k data (Hk : 0 < k), 
      exists lst, Contents_kdtree (build_kdtree k data) lst /\
                  Permutation data lst.

(** KNN Search **)

Lemma knn_search_build_size_gen :
    forall tree bb K k data query pq lst,
    priq _ _ pq ->
    Abs _ _ pq lst ->
    Contents_kdtree tree data ->
    size _ _ pq <= K ->
    size datapt (sum_dist query) (knn K k tree bb query pq) 
        = (min K ((length data) + (size _ _ pq))).

Lemma knn_search_build_size :
    forall K k data query,
    0 < k ->
    size datapt (sum_dist query) 
        (knn K k (build_kdtree k data)
                 (mk_bbox (repeat None k) (repeat None k)) 
                 query (empty _ _)) 
        = min K (length data).

Lemma perm_split_rearrange_all_in_leb :
    forall A key lst (a:list A) b c d
        (Hab : Permutation (a ++ b) lst)
        (Hde : Permutation (c ++ d) lst)
        (Hleb : @all_in_leb A key c d),
        exists a1 a2 b1 b2, Permutation (a1 ++ a2) a /\
                            Permutation (b1 ++ b2) b /\
                            Permutation (a1 ++ b1) c /\
                            Permutation (a2 ++ b2) d /\
                            all_in_leb key a1 a2 /\
                            all_in_leb key b1 b2.

Lemma knn_full_relate_gen :
    forall (K k : nat) (tree : kdtree) (bb : bbox) 
           (query : list nat)
           (pq : MaxPQ.priqueue datapt (sum_dist query))
           (lst data result : list datapt),
    0 < K ->
    kdtree_bounded tree bb = true ->
    Contents_kdtree tree data ->
    MaxPQ.priq datapt (sum_dist query) pq ->
    MaxPQ.Abs datapt (sum_dist query) pq lst ->
    MaxPQ.priq datapt (sum_dist query) (knn K k tree bb query pq) ->
    MaxPQ.Abs datapt (sum_dist query)
                                (knn K k tree bb query pq) result ->
    exists lstSm lstLg dataSm dataLg leftover : list datapt,
         Permutation (lstSm ++ lstLg) lst /\
         Permutation (dataSm ++ dataLg) data /\
         Permutation (lstSm ++ dataSm) result /\
         Permutation (lstLg ++ dataLg) leftover /\
         (MaxPQ.size datapt (sum_dist query) 
                            (knn K k tree bb query pq) < K ->
          leftover = []) /\ 
         all_in_leb (sum_dist query) result leftover
        
Lemma knn_full_relate :
    forall K k data query,
    0 < K ->
    0 < length data ->
    0 < k ->
    (forall v' : datapt, In v' data -> length v' = k) ->
    forall pq, 
    pq = (knn K k (build_kdtree k data) 
                  (mk_bbox (repeat None k) (repeat None k)) 
                  query (empty datapt (sum_dist query))) ->
    exists result leftover : list datapt,
        Permutation data (result ++ leftover) /\
        Abs datapt (sum_dist query) pq result /\
        all_in_leb (sum_dist query) result leftover.

Theorem knn_search_build_kdtree_correct :
    forall (K:nat) (k : nat) (data : list datapt),
    0 < K ->
    0 < length data ->
    0 < k ->
    (forall v' : datapt, In v' data -> length v' = k) ->
    forall tree query result,
    tree = (build_kdtree k data) ->
    knn_search K k tree query = result ->
    exists leftover, 
        length result = min K (length data)
        /\ Permutation data (result ++ leftover)
        /\ all_in_leb (sum_dist query) result leftover.

































\end{Verbatim}

\end{document}